\documentclass[conference]{IEEEtran}
\usepackage[dvips]{graphicx}
\usepackage{amsmath}
\usepackage{amsthm}
\usepackage[flushleft]{threeparttable}
\usepackage{array}
\usepackage{booktabs}
\usepackage{bbm}
\newcommand{\PreserveBackslash}[1]{\let\temp=\\#1\let\\=\temp}
\newcolumntype{C}[1]{>{\PreserveBackslash\centering}p{#1}}
\newcolumntype{R}[1]{>{\PreserveBackslash\raggedleft}p{#1}}
\newcolumntype{L}[1]{>{\PreserveBackslash\raggedright}p{#1}}

\usepackage{bm}
\usepackage{graphicx,subfigure}
\usepackage{epstopdf}
\usepackage{algorithm}
\usepackage{cite}
\usepackage{amsmath}
\usepackage{amssymb}
\usepackage{psfrag}
\usepackage{algpseudocode}
\usepackage{empheq}
\usepackage{latexsym, amsmath, subfigure, color, amsfonts, amssymb,graphicx}
\usepackage{wrapfig}
\usepackage{enumitem}

\usepackage{bbm, dsfont}
\newtheorem{Theorem}{Theorem}

\newtheorem{Definition}{Definition}

\newtheorem{Remark}{Remark}
\IEEEoverridecommandlockouts
\usepackage{tabularx}

\usepackage[T1]{fontenc}
\usepackage{etoolbox}

\makeatletter
\def\hlinewd#1{%
\noalign{\ifnum0=`}\fi\hrule \@height #1 %
\futurelet\reserved@a\@xhline}
\patchcmd{\maketitle}{\@fnsymbol}{\@alph}{}{}  
\makeatother
\def\hlinewd#1{%
\noalign{\ifnum0=`}\fi\hrule \@height #1 %
\futurelet\reserved@a\@xhline}
\patchcmd{\maketitle}{\@fnsymbol}{\@alph}{}{}  
\makeatother

\title{Centralized Coded Caching of Correlated Contents}

\author{Qianqian Yang and Deniz G\"{u}nd\"{u}z\\
\IEEEauthorblockA{Information Processing and Communications Lab\\ Department of Electrical and Electronic Engineering\\ Imperial College London
}
}

\date{}

\begin{document}

\maketitle
\begin{abstract}
Coded caching and delivery is studied taking into account the correlations among the contents in the library. Correlations are modeled as common parts shared by multiple contents; that is, each file in the database is composed of a group of subfiles, where each subfile is shared by a different subset of files. The number of files that include a certain subfile is defined as the \textit{level of commonness} of this subfile. First, a correlation-aware \textit{uncoded} caching scheme is proposed, and it is shown that the optimal placement for this scheme gives priority to the subfiles with the highest levels of commonness. Then a correlation-aware \textit{coded} caching scheme is presented, and the cache capacity allocated to subfiles with different levels of commonness is optimized in order to minimize the delivery rate. The proposed correlation-aware coded caching scheme is shown to remarkably outperform state-of-the-art correlation-ignorant solutions, indicating the benefits of exploiting content correlations in coded caching and delivery in networks.
\end{abstract}

\section{Introduction}\label{intro}
In \textit{proactive caching}, popular contents are stored in user devices during off-peak traffic periods even before they are requested by the users ~\cite{GregoryDtoD, Bastug:CM:14, samuel2017}. Proactive caching is considered as a promising solution for the recent explosive growth of wireless data traffic, and can alleviate both the network congestion and the latency during peak traffic periods (see \cite{GregoryDtoD, Bastug:CM:14, MaddahAliCentralized,samuel2017}, and references therein). 

Proactive caching typically takes place in two phases: the first phase takes place during off-peak traffic periods, when users' caches are filled as a function of the whole library of files, referred to as the \textit{placement phase}; while the second, \textit{delivery phase}, takes place during the peak traffic period when the users' demands are revealed and satisfied simultaneously. In contrast to traditional uncoded caching schemes, which simply employ orthogonal unicast transmissions during the \textit{delivery phase}, recently proposed \textit{coded caching}\cite{MaddahAliCentralized} creates coded multicasting opportunities, significantly reducing the amount of data that needs to be delivered to the users to satisfy their demands, even when these demands are distinct. Coded caching benefits from the aggregate cache capacity across the network, rather than local cache capacities as in conventional uncoded caching \cite{MaddahAliCentralized}. This significant improvement has motivated intense research interest on coded caching in recent years \cite{MaddahAliDecentralized,MohammadQianDenizITW,MohammadDenizTCom,VilardeboCodedCaching,JiArXivNonuniform,NiesenNonuniform,PedarsaniOnlineCaching,yang2016coded}. Some works follow the simplified model proposed in \cite{MaddahAliCentralized}, and aim to improve the fundamental limits of caching \cite{MohammadDenizTCom, MohammadQianDenizITW,yu2017exact}; others consider more realistic settings, such as decentralized caching \cite{MaddahAliDecentralized}, nonuniform popularities across files  \cite{NiesenNonuniform, JiArXivNonuniform}, audience retention rate aware caching \cite{yang2017audience}, or heterogeneous quality of service requirements\cite{yang2016coded}.

An important feature of video contents, which is the main source of the recent explosive traffic growth, is that, there may be significant overlaps among different files, e.g., the recordings of the same event from different angles and different cameras, or even the frames of the same scene in the same video. In \cite{hassanzadeh2016correlation}, Hassanzadeh et al. propose a correlation-aware caching scheme, which groups the contents in the library into two sets according to their correlations as well as popularity, where each file in second set is compressed with respect to a file in the first set. This scheme is shown to outperform correlation-ignorant caching schemes. A more information theoretic formulation for caching of correlated sources is considered in \cite{timo2016rate}, focusing on a special scenario with two receivers and one cache. A similar information-theoretic analysis is carried out in \cite{hassanzadeh:Arxiv:17} for two files and two receivers, each with its own cache. 

In this paper, we consider a server with a library of $N$ correlated files, serving $K$ users equipped with local caches. Different from \cite{hassanzadeh2016correlation}, which only exploits a fixed level of common information among the files, we consider a more general model in order to fully exploit the potential correlations among different subsets of files. We model each file in the server to be composed of a group of subfiles, such that each subfile is shared by a different subset of files. The number of files to which each subfile belongs is defined as its \textit{level of commonness}. Equivalently, in our model, any subset of files $\mathcal{S}$ share a common part that is independent of the rest of the library, and shared exclusively by the files in $\mathcal{S}$.

We first propose a correlation-aware \textit{uncoded} caching scheme, and show that the optimal placement for this scheme is achieved by giving priority to the subfiles with the highest levels of commonness in the placement phase. We then propose a correlation-aware \textit{coded} caching scheme, and derive a closed-form expression of the achievable delivery rate, based on which the cache capacity is optimally allocated to the subfiles according to their level of commonness. Then, we compare the performance of the proposed correlation-aware schemes with those that ignore the correlation, and the cut-set bound; and show that, exploiting file correlations in coded caching can significantly reduce the delivery rate.



\textit{Notations:} The set of integers $\left\{ i, ..., j \right\}$, where $i \le j$, is denoted by $\left[ i:j \right]$, particularly, $\left\{1, ..., j \right\}$ is denoted by $\left[j \right]$. For sets $\mathcal{A}$ and $\mathcal{B}$, we define $\mathcal{A} \backslash \mathcal{B}\triangleq\{x: x \in \mathcal{A}, x\notin \mathcal{B}\}$, and $\left| \mathcal{A} \right|$ denotes the cardinality of $\mathcal{A}$. $\binom{j}{i}$ represents the binomial coefficient if $j\geq i$; otherwise, $\binom{j}{i}=0$. For event $E$, $\mathbbm{1}\{E\}=1$ if $E$ is true; and $\mathbbm{1}\{E\}=0$, otherwise.

\section{System Model}\label{sys}
We consider a server with a database of $N$ correlated files, $W_1, ..., W_N$, where each file consists of $2^{N-1}$ independent subfiles, e.g., $W_i=\bigcup\limits_{\substack{\mathcal{S}\subset [N]\\ i \in \mathcal{S}}}\overline{W}_{\mathcal{S}}$, $\forall i\in [N]$. Here, $\overline{W}_{\mathcal{S}}$ denotes the subfile shared exclusively by the subset of files $\{W_i: i \in \mathcal{S}\}$. For simplicity, we assume that for $\mathcal{S}\subset [N]$, $|\overline{W}_{\mathcal{S}}|=F_l$, if $|\mathcal{S}|=l$, i.e., the common subfiles shared exclusively by $l$ files are of the same size of $F_l$ bits. Let $\mathbf{F} \triangleq (F_1, \ldots, F_N)$. As a result, each file in the library is also of the same size of $F$ bits, given by 
\begin{equation}
F=\sum\limits_{l=1}^N \binom{N-1}{l-1}F_l.
\end{equation}

For $\mathcal{S}\subset [N]$, $|\mathcal{S}|=l$, we say that the subfiles $\overline{W}_{\mathcal{S}}$ have a commonness level of $l$. For example, $\overline{W}_{\{1, 2, 3\}}$ and $\overline{W}_{\{3, 4, 5\}}$ both have level $3$ commonness. For brevity, we refer to all the subfiles with level $l$ commonness as $l$-subfiles, $l=1, ..., N$.  
We consider $K$ users connected to the server through a shared, error-free link, each equipped with a cache of size $MF$ bits. 

We consider centralized caching; that is, the server has the knowledge of the active users during the placement phase, though not the knowledge of their demands. Centralized caching allows the server to fill the user caches in a coordinated manner. After the placement phase, each user requests a single file from the library, where $d_k \in [N]$ denotes user $k$'s request, $k\in [K]$. All the requests are satisfied simultaneously over the error-free shared link.

An $(\mathbf{F}, M, R)$ caching code for this system consists of:
\begin{itemize}
\item \textbf{$K$ caching functions} $f_{k}$, $k \in [K]$,
\begin{equation}
  f_{k}: \underbrace{[2^F] \times  \cdots \times [2^F]}\limits_{N~\text{files}}\rightarrow [2^{MF}],
\end{equation}
such that the contents of user $k$'s cache at the end of the placement phase, denoted by $Z_k$, is given by $Z_k=f_{k}(\{W_i\}_{i=1}^N)$;
\item \textbf{a delivery function} $g$, 
\begin{equation}
g: \underbrace{[2^F] \times  \cdots \times [2^F]}\limits_{N~\text{files}} \times \mathbf{D} \rightarrow [ 2^{RF}],
\end{equation}
where $\mathbf{D}\triangleq (d_1, ..., d_K)$, such that a single message of $RF$ bits, $X_{\mathbf{D}}=g((W_1, ..., W_N), \mathbf{D})$, is sent by the server over the shared link according to users' demands;  
\item \textbf{$K$ decoding functions} $h_k$, $k \in [K]$,
\begin{equation}
h_k:  \mathbf{D} \times [2^{MF}]\times [2^{RF}] \rightarrow [2^{F}],
\end{equation}
where $\hat{W}_{d_k}=h_k (\mathbf{D}, Z_k, X_{\mathbf{D}})$, is the reconstruction of $W_{d_k}$ at user $k$.
\end{itemize}

\begin{Definition}
A user cache capacity-delivery rate pair $(M, R)$ is \textit{achievable} for a system described above, if there exists a sequence of $(\mathbf{F}, M, R)$ codes such that for any demand realization $\mathbf{D} \subset [N]^K$, 
\begin{equation}
\lim_{F_1, \ldots, F_N \rightarrow \infty} \Pr \left\{\bigcup\limits_{k\in [K]} \Big\{\hat{W}_{d_k}\neq W_{d_k}\Big\}\right\}=0.
\end{equation}
\end{Definition}
For a system with $N$ files and $K$ users, our goal is to characterize the minimum achievable rate $R$ as a function of the user cache capacity $M$, i.e., $R^*(M)\triangleq \inf\{R: (M, R) \mbox{ is achievable}\}$.

\section{Correlation-aware Uncoded Caching and Delivery (CAUC) Scheme}\label{s:Uncoded}

We first present an uncoded caching and delivery scheme exploiting the correlation among files, referred to as CAUC. 

\subsubsection{Placement phase} Each user caches the same $p_lF_l$ bits from each $l$-subfile, where $0 \leq p_l\leq 1$, $l\in [N]$, such that
\begin{equation}\label{cachecapacity}
MF=\sum\limits_{l=1}^N \binom{N}{l}p_lF_l,
\end{equation}
which meets the limitation of the cache capacities. We refer to $\mathbf{P} \triangleq (p_{1}, ..., p_N)$ as the \textit{cache allocation vector}, which will be specified in the sequel.

\subsubsection{Delivery phase} The server delivers the remaining bits of each requested subfile that have not been cached by the users, i.e., $\overline{W}_{\mathcal{S}}$ for which $\sum\limits_{k=1}^{K}\mathbbm{1}\{d_k \in \mathcal{S}\} \geq 1$.    

In the worst case, when the demand combination is the most distinct, i.e., users request distinct files for the case $N\geq K$, or each file is requested by at least one user for the case $N <K$, the delivery rate is given by \begin{equation}\label{uncodedrate}
R_{CAUC}(\mathbf{P})=\sum\limits_{l=1}^{N}(1-p_l)F_l{\small\left(\binom{N}{l}-\binom{\min\{N-K, 0\}}{l}\right)}.
\end{equation}
The optimal $\mathbf{P}^*$ can be derived by solving the following optimization problem
\begin{equation}
\begin{aligned}
&\min~~~R_{CAUC}(\mathbf{P})\\
&\text{such that}~~~\sum\limits_{l=1}^N \binom{N}{l}p_lF_l\leq MF,
\end{aligned}
\end{equation} 
which, straightforwardly, leads to: $p^*_l=1$, if $C(l) \leq MF$; $p^*_l=\frac{MF-C(l+1)}{\binom{N}{l}F_l}$, if $C(l+1) < MF< C(l)$; and $p^*_l=0$, otherwise; where we have defined $C(l) \triangleq \sum\limits_{i=l}^N \binom{N}{i}F_i$, for $l\in [N]$. We remark that the optimal cache allocation gives priority to the subfiles with the highest level of commonness. 

\section{Correlation-aware Coded Caching and Delivery (CACC) Scheme}\label{s:Coded}

In this section, we present a correlation-aware coded caching and delivery scheme, referred to as CACC. Similarly to the CAUC scheme, we allocate different cache capacities to subfiles of different levels of commonness, again specified by the cache allocation vector, $\mathbf{P}=(p_{1}, ..., p_N)$, which satisfies the constraint in \eqref{cachecapacity}, such that each user caches $p_lF_l$ bits from each $l$-subfile, $l\in [N]$. In the following, we first present how coded caching and delivery of the subfiles with the same level of commonness is carried out, and then specify the allocation of cache capacity.

\subsection{Coded Caching and Delivery of $l$-subfiles}

Here, for a given cache allocation vector $\mathbf{P}$, we present the coded caching and delivery of $l$-subfiles, $l\in[N]$. We define $t_l\triangleq Kp_l$, $0\leq t_l \leq K$. If $t_l=0$, users do not cache the $l$-subfiles at all, while if $t_l=K$, each user stores all the $l$-subfiles in its cache. In the following, we focus on the cases where $t_l \in [K-1]$.

\subsubsection{Placement Phase}
We employ the prefetching scheme proposed by~\cite{MaddahAliCentralized} for the subfiles rather than the files themselves: each $l$-subfile is partitioned into $\binom{K}{t_l}$ disjoint parts, each with approximately the same size of $F_l/\binom{K}{t_l}$ bits. We label these $\binom{K}{t_l}$ disjoint parts of each $l$-subfile $\overline{W}_{\mathcal{S}}$ by $\overline{W}_{\mathcal{S}}^{\mathcal{A}}$, where $|\mathcal{A}|=t_l,~ \mathcal{A}\subset [K]$; that is, we have $\overline{W}_{\mathcal{S}}= \bigcup_{\mathcal{A}: |\mathcal{A}|=t_l,~ \mathcal{A}\subset [K]}  \overline{W}_{\mathcal{S}}^{\mathcal{A}}$. Each of these parts, $\overline{W}_{\mathcal{S}}^{\mathcal{A}}$, is placed into the cache of user $k$ if $k \in \mathcal{A}$. Thus, each user caches a total of $\binom{K-1}{t_l-1}$ disjoint parts of each $l$-subfile with a total size of $t_lF_l/K$ bits, which sums up to $p_lF_l$ bits. 

\subsubsection{Delivery Phase}
We first focus on the case when $N\leq K$. There are a total of $\binom{N}{l}$ $l$-subfiles. We denote the set of these $l$-subfiles by $\mathcal{W}^l=\left\{\overline{W}_{\mathcal{S}}: \mathcal{S} \subset [N], |\mathcal{S}|=l\right\}$. Each user requires a total of $\binom{N-1}{l-1}$ $l$-subfiles, i.e., user $k$ needs to recover subfiles in $\left\{\overline{W}_{\mathcal{S}}: \mathcal{S} \subset [N], |\mathcal{S}|=l, d_k\in \mathcal{S}\right\}$, $\forall k\in [K]$. For each user, we can regard these $\binom{N-1}{l-1}$ $l$-subfiles as $\binom{N-1}{l-1}$ distinct demands. Our delivery scheme for the $l$-subfiles operates in $\binom{N-1}{l-1}$ steps, and satisfies one demand of each user at each step.

We define $\mathbf{C}_j\triangleq(c_{1j}, ..., c_{Nj})$, where $c_{ij} \in \{\mathcal{S}: \mathcal{S} \subset [N], |\mathcal{S}|=l, i\in \mathcal{S}\}$, $\forall i\in [N]$, $j\in [\binom{N-1}{l-1}]$, which specifies which subfile should be delivered in the $j$th step of the delivery phase. $\mathbf{C}_j$ is generated by Algorithm \ref{groupingscheme} by setting $\mathcal{R}=[N]$ and $\overline{\mathcal{R}}=\emptyset$. Note that these vectors are generated independently of the number of users or their demands.

\theoremstyle{definition}
\newtheorem{exmp}{Example}

\begin{exmp}
Consider $N=5$ and $l=2$. From Algorithm \ref{groupingscheme} we obtain:
\begin{align*}
&\mathbf{C_1}=(\{1,2\}, \{1,2\}, \{3,4\}, \{3,4\}, \{1,5\});\\  &\mathbf{C}_2=(\{1,5\}, \{2,3\},\{2,3\}, \{4,5\},\{4, 5\});\\
&\mathbf{C}_3=(\{1,3\}, \{2,5\}, \{1,3\}, \{2,4\},  \{2,5\});\\
&\mathbf{C}_4=(\{1,4\}, \{2,4\}, \{3,5\}, \{1,4\}, \{3, 5\}).
\end{align*}
This means, for example, that, in the first step, subfiles $W_{12}, W_{34},$ and $W_{15}$ will be delivered (if there is a user requesting them). 
\end{exmp}

We denote by $d^j_k\triangleq c_{d_kj}$ the demand of user $k$ to be satisfied in the $j$th step, i.e., user $k$ recovers $\overline{W}_{d^j_k}$ after the $j$th step. We emphasize that $\bigcup\limits_{j=1}^{\binom{N-1}{l-1}}\overline{W}_{c_{ij}}=\{\overline{W}_{\mathcal{S}}: \mathcal{S} \in [N], |\mathcal{S}|=l, i\in \mathcal{S}\}$, $\forall i \in [N]$; that is all the required $l$-subfiles will be recovered by each user after step $\binom{N-1}{l-1}$ for any demand combination.   

\begin{algorithm}[t]
\caption{Generate $\mathbf{C}_j$, $j\in[\binom{|\mathcal{R}|-1}{l-\mathcal{\overline{R}}-1}]$}
\label{groupingscheme}
\begin{algorithmic}[1]
\Statex
\Procedure { Assignment}{}
\State{ $\mathcal{W}_t \leftarrow  \mathcal{W}^l$, $j \leftarrow 1$, $W_{last} \leftarrow  \emptyset$, $c_{last} \leftarrow  \emptyset$}
\While{$\mathcal{W}_t\not= \emptyset$}
\State{$c_{t} \leftarrow \mathcal{R}$}
\While{$c_{t}\not= \emptyset$}
\If{$|c_{t}|\geq l-|\overline{\mathcal{R}}|$}
\If{$c_{last}=\emptyset$}
\State{Randomly select one $l$-subfile $\overline{W}_{\mathcal{S}}$ from $\mathcal{W}_t$ such that $\mathcal{S}\setminus \overline{\mathcal{R}} \subset c_{t}$, remove $\overline{W}_{\mathcal{S}}$ from $\mathcal{W}_t$, and $c_{t} \leftarrow  c_{t}\setminus \mathcal{S}$}
\For{$i \in \mathcal{S}\setminus \overline{\mathcal{R}}$}
\State{$c_{ij} \leftarrow  W_{last}$}
\EndFor
\Else 
\For{$i \in c_{last}\setminus \overline{\mathcal{R}}$}
\State{$c_{ij} \leftarrow  \mathcal{S}$}
\EndFor
\State{$c_{t} \leftarrow  c_{t}\setminus c_{last}$, $W_{last} \leftarrow  -1$, and $c_{last} \leftarrow \emptyset$}
\EndIf
\Else
\State{Randomly select one $l$-subfile $\overline{W}_{\mathcal{S}}$ from $\mathcal{W}_t$ such that $c_{t}\subset \mathcal{S}$}
\For{$i \in c_{t}$}
\State{$c_{ij} \leftarrow  \mathcal{S}$}
\EndFor
\For{$i \in \overline{\mathcal{R}}$}
\State{$c_{ij} \leftarrow  \emptyset$}
\EndFor
\State{$W_{last} \leftarrow  \mathcal{S}$, $c_{last} \leftarrow  \mathcal{S}\setminus c_{t}$, $c_{t} \leftarrow \emptyset$, and $j \leftarrow  j+1$}
\EndIf
\EndWhile
\EndWhile
\EndProcedure
\end{algorithmic}
\end{algorithm}

\begin{exmp}\label{ex:2}
Consider $N=K=5$ and $l=2$ as in Example 1. Consider distinct demands, i.e., $\mathbf{D} = \{1, 2, 3, 4, 5\}$. Thus, based on $\mathbf{C}_1$, we have $d^1_1=d^1_2=\{1, 2\}$, $d^1_3=d^1_4=\{3, 4\}$, and $d^1_5=\{1, 5\}$; that is, at the end of the first step, users 1 and 2 should recover $W_{12}$, users 3 and 4 should recover $W_{34}$, while user 5 should recover $W_{15}$.
\end{exmp}

\begin{exmp}\label{ex:3}
With the same setting as in Example 2, consider now a non-distinct demand combination $\mathbf{D}=\{1, 1, 1, 3, 4\}$. Based on $\mathbf{C}_1$, we have $d^1_1=d^1_2$ $=d^1_3$ $=\{1, 2\}$, and $d^1_4$ $=d^1_5=\{3, 4\}$; that is, at the end of the first step users 1, 2 and 3 should recover $W_{12}$, while users 4 and 5 should recover $W_{34}$.
\end{exmp}

Based on the delivery scheme proposed in \cite{yu2017exact}, we present our coded transmission scheme in Algorithm \ref{deliveryscheme} according to $\mathbf{C}_j$, $j\in [\binom{N-1}{l-1}]$, where $\overline{\mathcal{R}}=\emptyset$, $\mathcal{R}=[N]$. In Algorithm \ref{deliveryscheme}, we define $A_j$ as the number of distinct $d^j_k$ for each $j \in [\binom{N-1}{l-1}]$. We note that, among the CODED DELIVERY and RANDOM DELIVERY procedures of Algorithm.~\ref{deliveryscheme}, the one that requires a smaller delivery rate is performed.

\begin{algorithm}[t]
\caption{Coded transmission based on $\mathbf{C}_j$}
\label{deliveryscheme}
\begin{algorithmic}[1]
\Statex
\Procedure{Coded Delivery}{}
\For{$k=1, ..., K$}
\State{$d^j_k \leftarrow c_{d_kj}$}
\EndFor
\State{$\mathcal{U}_j \leftarrow$ Any subset of $A_j$ users with distinct $d^j_k$}
\For{$\mathcal{V}\subset [K]: |\mathcal{V}|=t_l+1, \sum\limits_{k \in \mathcal{U}_j}\mathbbm{1}\{k \in \mathcal{V}\} \geq 1$}
\State{Send ${\bigoplus}_{k\in \mathcal{V}}\overline{W}^{\mathcal{V}\setminus \{k\}}_{d^j_k}$.}
\EndFor
\EndProcedure
\Statex
\Procedure {Random Delivery}{}
\For{$\mathcal{S}\subset \mathcal{R}: |\mathcal{S}|=l-|\overline{\mathcal{R}}|$}
\State {Server sends enough random linear combinations of the bits of $l$-subfile $\overline{W}_{\mathcal{S}\cup \overline{\mathcal{R}}}$ to enable the users demanding it to decode it.}
\EndFor
\EndProcedure
\end{algorithmic}
\end{algorithm}

\textbf{Example \ref{ex:2} - continued.} 
In Example \ref{ex:2}, assume that $t_l=1$, i.e., each $l$-subfile is divided into $K$ disjoint parts of equal size, and each disjoint part is cached exactly by one user. Based on $\mathbf{D}$, we have $A_1=3$. Assume that $\mathcal{U}_1=\{1, 3, 5\}$. Then, the server sends $\overline{W}_{\{1, 2\}}^{\{1\}}\bigoplus\overline{W}_{\{1, 2\}}^{\{2\}}$, $\overline{W}_{\{3, 4\}}^{\{1\}}\bigoplus\overline{W}_{\{1, 2\}}^{\{3\}}$,  $\overline{W}_{\{3, 4\}}^{\{1\}}\bigoplus\overline{W}_{\{1, 2\}}^{\{4\}}$,  $\overline{W}_{\{1, 5\}}^{\{1\}}\bigoplus\overline{W}_{\{1, 2\}}^{\{5\}}$, $\overline{W}_{\{3, 4\}}^{\{2\}}\bigoplus\overline{W}_{\{1, 2\}}^{\{3\}}$, $\overline{W}_{\{3, 4\}}^{\{3\}}\bigoplus\overline{W}_{\{3, 4\}}^{\{4\}}$, $\overline{W}_{\{3, 4\}}^{\{5\}}\bigoplus\overline{W}_{\{1, 5\}}^{\{3\}}$, $\overline{W}_{\{1, 5\}}^{\{2\}}\bigoplus\overline{W}_{\{1, 2\}}^{\{5\}}$,  $\overline{W}_{\{1, 5\}}^{\{4\}}\bigoplus\overline{W}_{\{3, 4\}}^{\{5\}}$. By receiving these coded bits, users $1$ and $2$ can recover $\overline{W}_{\{1, 2\}}$ together with the contents of their own caches. Similarly, users $3$ and $4$ can recover $\overline{W}_{\{3, 4\}}$, while user $5$ recovers $\overline{W}_{\{1, 5\}}$. In the same manner, by coded transmission based on $\mathbf{C}_2$, user $1$ can recover $\overline{W}_{\{1, 5\}}$, users $2$ and $3$ recover $\overline{W}_{\{2, 3\}}$, and users $4$ and $5$ recover  $\overline{W}_{\{4, 5\}}$ in the second step. After four delivery steps based on $\mathbf{C}_1$, \ldots, $\mathbf{C}_4$ each user decodes all the $2$-subfiles of their requests. The total number of bits delivered in these four steps is $36F_2/5$.     

\textbf{Example \ref{ex:3} - continued.} 
Assume again that $t=1$. Based on $\mathbf{D}$, we have $A_1=2$, and let $\mathcal{U}_1=\{1, 4\}$. Then, the server sends $\overline{W}_{\{1, 2\}}^{\{1\}}\bigoplus\overline{W}_{\{1, 2\}}^{\{2\}}$, $\overline{W}_{\{1, 2\}}^{\{1\}}\bigoplus\overline{W}_{\{1, 2\}}^{\{3\}}$,  $\overline{W}_{\{3, 4\}}^{\{1\}}\bigoplus\overline{W}_{\{1, 2\}}^{\{4\}}$,  $\overline{W}_{\{3, 4\}}^{\{1\}}\bigoplus\overline{W}_{\{1, 2\}}^{\{5\}}$, $\overline{W}_{\{3, 4\}}^{\{2\}}\bigoplus\overline{W}_{\{1, 2\}}^{\{4\}}$, $\overline{W}_{\{3, 4\}}^{\{3\}}\bigoplus\overline{W}_{\{1, 2\}}^{\{4\}}$, $\overline{W}_{\{3, 4\}}^{\{5\}}\bigoplus\overline{W}_{\{3, 4\}}^{\{4\}}$, such that users $1, 2$ and $3$ can recover $\overline{W}_{\{1, 2\}}$, while users $4$ and $5$ can recover $\overline{W}_{\{3, 4\}}$. Based on $\mathcal{C}_2$, the server sends $\overline{W}_{\{1, 5\}}^{\{1\}}\bigoplus\overline{W}_{\{1, 5\}}^{\{2\}}$, $\overline{W}_{\{1, 5\}}^{\{1\}}\bigoplus\overline{W}_{\{1, 5\}}^{\{3\}}$,  $\overline{W}_{\{2, 3\}}^{\{1\}}\bigoplus\overline{W}_{\{1, 5\}}^{\{4\}}$,  $\overline{W}_{\{4, 5\}}^{\{1\}}\bigoplus\overline{W}_{\{1, 5\}}^{\{5\}}$, $\overline{W}_{\{2, 3\}}^{\{2\}}\bigoplus\overline{W}_{\{1, 5\}}^{\{4\}}$, $\overline{W}_{\{2, 5\}}^{\{3\}}\bigoplus\overline{W}_{\{1, 5\}}^{\{4\}}$, $\overline{W}_{\{2, 3\}}^{\{5\}}\bigoplus\overline{W}_{\{4, 5\}}^{\{4\}}$, $\overline{W}_{\{4, 5\}}^{\{2\}}\bigoplus\overline{W}_{\{1, 5\}}^{\{5\}}$, $\overline{W}_{\{4, 5\}}^{\{3\}}\bigoplus\overline{W}_{\{1, 5\}}^{\{5\}}$, such that users $1, 2$ and $3$ can recover $\overline{W}_{\{1, 5\}}$, while user $4$ and user $5$ can recover $\overline{W}_{\{2, 3\}}$ and $\overline{W}_{\{4, 5\}}$, respectively. Based on $\mathbf{C}_3$, the server sends $\overline{W}_{\{1, 3\}}^{\{1\}}\bigoplus\overline{W}_{\{1, 3\}}^{\{2\}}$, $\overline{W}_{\{1, 3\}}^{\{1\}}\bigoplus\overline{W}_{\{1, 3\}}^{\{3\}}$,  $\overline{W}_{\{1, 3\}}^{\{1\}}\bigoplus\overline{W}_{\{1, 3\}}^{\{4\}}$,  $\overline{W}_{\{2, 4\}}^{\{1\}}\bigoplus\overline{W}_{\{1, 3\}}^{\{5\}}$, $\overline{W}_{\{2, 4\}}^{\{2\}}\bigoplus\overline{W}_{\{1, 3\}}^{\{5\}}$, $\overline{W}_{\{2, 4\}}^{\{3\}}\bigoplus\overline{W}_{\{1, 3\}}^{\{5\}}$, $\overline{W}_{\{2, 4\}}^{\{4\}}\bigoplus\overline{W}_{\{1, 3\}}^{\{5\}}$, based on which users $1$,$2$,$3$ and $4$ can recover $\overline{W}_{\{1, 3\}}$, while user $5$ recovers $\overline{W}_{\{2, 4\}}$. Finally, based on $\mathbf{C}_4$, the server sends $\overline{W}_{\{1, 4\}}^{\{1\}}\bigoplus\overline{W}_{\{1, 4\}}^{\{2\}}$, $\overline{W}_{\{1, 4\}}^{\{1\}}\bigoplus\overline{W}_{\{1, 4\}}^{\{3\}}$,  $\overline{W}_{\{1, 4\}}^{\{1\}}\bigoplus\overline{W}_{\{1, 4\}}^{\{5\}}$,  $\overline{W}_{\{3, 5\}}^{\{1\}}\bigoplus\overline{W}_{\{1, 4\}}^{\{4\}}$, $\overline{W}_{\{3, 5\}}^{\{2\}}\bigoplus\overline{W}_{\{1, 4\}}^{\{4\}}$, $\overline{W}_{\{3, 5\}}^{\{3\}}\bigoplus\overline{W}_{\{1, 4\}}^{\{4\}}$, $\overline{W}_{\{3, 5\}}^{\{5\}}\bigoplus\overline{W}_{\{1, 4\}}^{\{4\}}$, such that users $1, 2, 3$ and $4$ are able to recover $\overline{W}_{\{1, 4\}}$, while user $4$ recovers $\overline{W}_{\{3, 5\}}$. Thus, all the users are able to decode the $2$-subfiles they requested. The total number of bits delivered in this case is $30F_2/5$.        

Next, we consider the case $N>K$. We first select a subset of $K$ files $\mathcal{R}$, $\mathcal{R} \subset [N]$, such that $|\mathcal{R}|=K$, and $d_k \in \mathcal{R}$ for $k=1, ..., K$. For any subset $\overline{\mathcal{R}} \in [N]\setminus \mathcal{R}$, $|\overline{\mathcal{R}}|=s$, $s \in [\max\{l-K, 0\}: \min\{l-1, N-K\}\}]$,  Algorithm \ref{groupingscheme} is applied to a subset of $l$-subfiles $\mathcal{W}^l=\{\overline{W}_{\mathcal{S}\cup\overline{\mathcal{R}}}: \mathcal{S} \subset \mathcal{R}, |\mathcal{S}|=l-|\overline{\mathcal{R}}|\}$ to derive $\mathbf{C}_j$, $j\in[\binom{|\mathcal{R}|-1}{l-\mathcal{\overline{R}}-1}]$, based on which Algorithm \ref{deliveryscheme} is applied to enable each user to decode its demanded $l$-subfiles in $\mathcal{W}^l$. Therefore, each user can decode all the $l$-subfiles it is demanding.

\subsection{Achievable rate}
The following theorem presents the delivery rate achieved by the proposed coded caching and delivery scheme for any demand combination, for a given cache allocation vector $\mathbf{P}$. 

\begin{Theorem}\label{theorem:rate}
For the caching system described in Section~\ref{sys}, given a cache allocation vector $\mathbf{P}$, the following delivery rate is achievable  
\begin{equation}\label{averagerate}
R_{CACC}(\mathbf{P})=\sum\limits_{l=1}^N R_l(t_l), 
\end{equation}
where $t_l=p_lK$, and for $t_l \in [0: K]$,
\begin{equation}\label{Rleq}
R_l(t_l)= \min\{\alpha_l(t_l), m_l(t_l)\},
\end{equation}
and
\begin{equation}\label{ratecodeddelivery}
\begin{aligned}
&\alpha_l(t_l) \triangleq \sum\limits_{s=\max\{l-K, 0\}}^{\max\{\min\{l-1, N-K\}, 0\}}\binom{N-K}{s}\binom{\min\{N, K\}-1}{l-s-1}\cdot\\
&\left[\binom{K}{t_l+1}-\binom{\max\{K-\lceil \frac{\min\{N, K\}}{l-s} \rceil -1, 0\}}{t_l+1}\right]\frac{F_l}{F\binom{K}{t_l}},
\end{aligned}
\end{equation}
\begin{equation}
\begin{aligned}\label{varphi}
m_l(t_l) \triangleq \left(\binom{N}{l}-\binom{\min\{N-K, 0\}}{l}\right) (F_l-t_lF_l/K)/F.
\end{aligned}
\end{equation}
For $t_l \notin [0: K]$, $R_l(t_l)$ is given by the lower convex envelop of the above achievable points.
\begin{proof}
We show that $R_l(t_l)$ given above is achievable $t_l \in [0:K]$. The lower convex envelop of these integer points can then be achieved by memory sharing. For the case $N\leq K$, recall that the requested $l$-subfiles are delivered in $\binom{N-1}{l-1}$ steps based on $\mathbf{C}_j$, $j=[\binom{N-1}{l-1}]$, derived by Algorithm \ref{groupingscheme}. In each step, the server sends at most $\lceil N/l \rceil+1$ $l$-subfiles. Therefore, for any demand combination, the number of distinct $d^j_k$ based on $\mathbf{C}_j$, i.e., $A_j\leq \lceil N/l \rceil+1$, $n=1, ..., \binom{N-1}{l-1}$. Similar to the delivery scheme proposed in \cite{yu2017exact}, by the CODED DELIVERY procedure of Algorithm \ref{deliveryscheme}, the server broadcasts binary sums that help at least one user in $\mathcal{U}_j$ based on $\mathbf{C}_j$. The total number of such subsets of $t_1+1$ users that contain at least one user in $\mathcal{U}_j$ is given by $\binom{K}{t_l+1}-\binom{\max\{K-\lceil N/l \rceil -1, 0\}}{t_l+1}$. Hence, given $t_l \in \{1, ..., K-1\}$, the total number of bits sent by CODED DELIVERY procedure of Algorithm \ref{deliveryscheme} for the delivery of the $l$-subfiles is bounded by (normalized by $F$):
\footnotesize
\begin{equation}\label{rateneqkcodeddelivery}
R_l(t_l)\leq\binom{N-1}{l-1}\left[\binom{K}{t_l+1}-\binom{\max\{K-\lceil N/l \rceil -1, 0\}}{t_l+1}\right]\frac{F_l}{F\binom{K}{t_l}}
\end{equation}
\normalsize
The right hand side (RHS) of \ref{rateneqkcodeddelivery} is equal to \eqref{ratecodeddelivery} for $N\leq K$. Since each user caches $t_lF_l/K$ bits of each $l$-subfile, according to \cite[Appendix A]{MaddahAliDecentralized}, the number of bits sent by the RANDOM DELIVERY procedure is bounded by 
\footnotesize
\begin{equation}\label{rateneqkrandomdelivery}
R_l(t_l) \leq \binom{N}{l} (F_l-t_lF_l/K)/F. 
\end{equation}
\normalsize
The RHS equals to \eqref{varphi} for $N\leq K$. Hence, for $t_l \in \{1, ..., K-1\}$, we have proven $R_l(t_l)$ given in \eqref{Rleq} is achievable for $N\leq K$.

We then focus on the case where $N> K$. For any $ \max\{l-
K, 0\} \leq s\leq \min\{l-1, N-K\}$, there are a total of $\binom{N-K}{s}$ subsets $\overline{\mathcal{R}}$ such that $\overline{\mathcal{R}} \in [N]\setminus \mathcal{R}$, $|\overline{\mathcal{R}}|=s$. Following the similar analysis for the case where $N \leq K$, given $\mathcal{R}$ containing all the demanded files such that $|\mathcal{R}|=K$, and any $\overline{\mathcal{R}}$ such that $\overline{\mathcal{R}} \in [N]\setminus \mathcal{R}$, $|\overline{\mathcal{R}}|=s$, requested $l$-subfiles in $\mathcal{W}^l=\{\overline{W}_{\mathcal{S}\cup\overline{\mathcal{R}}}: \mathcal{S} \subset \mathcal{R}, |\mathcal{S}|=l-|\overline{\mathcal{R}}|\}$ are sent in $\binom{K-1}{l-s-1}$ step. At each step, there are at most $\lceil \frac{N}{l-s} \rceil+1$ $l$-subfiles to be sent. Therefore, with similar arguments, the total number of bits sent by CODED DELIVERY procedure of Algorithm \ref{deliveryscheme} in each step is bounded by $\binom{K-1}{l-s-1}\left(\binom{K}{t_l+1}-\binom{\max\{K-\lceil \frac{N}{l-s} \rceil -1, 0\}}{t_l+1}\right)\frac{F_l}{F\binom{K}{t_l}}$, while the number of bits send by the RANDOM DELIVERY procedure is bounded by $\binom{K}{l-s}(F_l-t_lF_l/K)/F$. By summing over all $\binom{N-K}{s}$ subsets $\overline{\mathcal{R}}$ for each $s \in [\max\{l-K, 0\}: \min\{l-1, N-K\}]$, we have 
\footnotesize
\begin{equation}\label{rateneqkcodeddelivery1}
\begin{aligned}
&R_l(t_l)\leq\sum\limits_{s=\max\{l-K, 0\}}^{\min\{l-1, N-K\}}\binom{N-K}{s}\binom{ K-1}{l-s-1}\cdot\\
&\left(\binom{K}{t_l+1}-\binom{\max\{K-\lceil \frac{N}{l-s} \rceil -1, 0\}}{t_l+1}\right)\frac{F_l}{F\binom{K}{t_l}},
\end{aligned}
\end{equation}
\normalsize
and, 
\footnotesize
\begin{equation}\label{rateneqkrandomdelivery1}
R_l(t_l) \leq \sum\limits_{s=\max\{l-K, 0\}}^{\min\{l-1, N-K\}}\binom{N-K}{s}\binom{K}{l-s} (F_l-t_lF_l/K)/F, 
\end{equation}
\normalsize
by which, we have proven the correctness of \eqref{Rleq} for the case $N>K$. 
\end{proof}
\end{Theorem}
\begin{Remark}
At each step of sending $l$-subfiles in $\mathcal{W}^l=\{\overline{W}_{\mathcal{S}\cup\overline{\mathcal{R}}}: \mathcal{S} \subset \mathcal{R}, |\mathcal{S}|=l-|\overline{\mathcal{R}}|\}$, there are sometimes $\lceil \frac{N}{l-s} \rceil+1$ and sometimes $\lceil \frac{N}{l-s} \rceil$ distinct demands, while when $N$ is a multiple of $l-s$, there are always $\frac{N}{l-s}$ distinct demands ($\mathcal{R}=[N]$, $\overline{\mathcal{R}}=\emptyset$, $s=0$, for the case $N\leq K$). To obtain a closed-form expression for the achievable delivery rate, we simply assume $\lceil \frac{N}{l-s} \rceil+1$ distinct demands at each step. Note that, the more the number of distinct demands at each step, the larger the delivery rate. Therefore $R_{CACC}(\mathbf{P})$ in \eqref{averagerate} is an upper bound on the actual achievable delivery rate of CACC. 
\end{Remark}
\subsection{Allocation of Cache Capacity}
We can further optimize the cache content distribution $\mathbf{P}$ by solving: 
\begin{subequations}\label{optimization}
\begin{align}
\min~ & R_{CACC}(\mathbf{P})\label{object1}\\
\mathrm{such~that~} & \sum\limits_{l=1}^N \binom{N}{l}p_lF_l\leq MF,\label{constrain}
\end{align}
\end{subequations}
where the objective is to minimize the achievable delivery rate under the cache capacity constraint. The problem in \eqref{optimization} can be solved numerically.

\begin{figure}[t]\label{fig:1}
\centering
\includegraphics[width=1.1\linewidth]{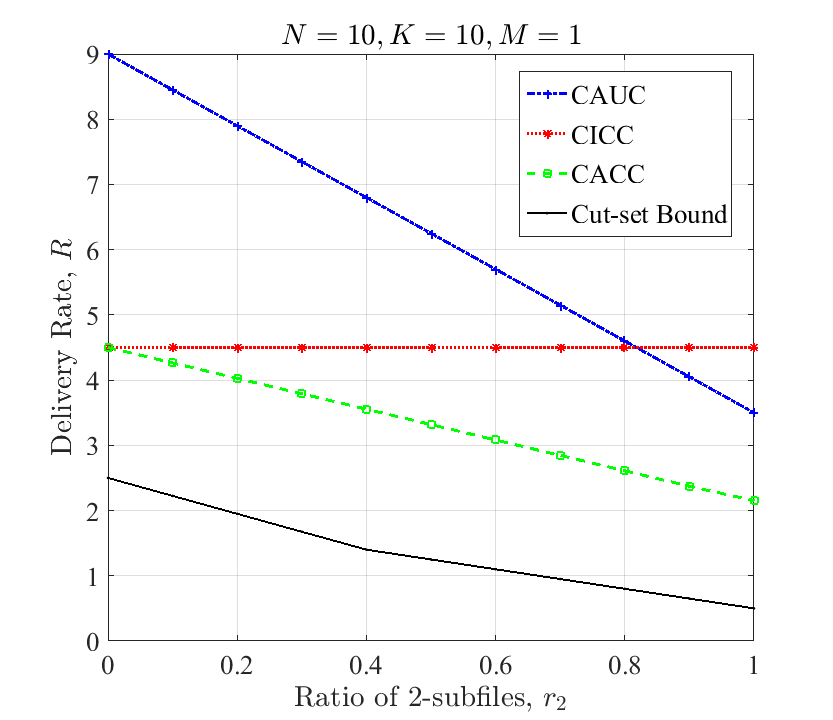}
\caption{Delivery rate ($R$) vs. ratio of $2$-subfiles ($r_2$),\newline $r_3=\cdots=r_{10}=0$.}
\end{figure}

\section{Lower Bound}\label{s:lower_bound}

In this section, we present a lower bound derived using cut-set arguments.
\begin{Theorem}\label{cutset}(Cut-set Bound)
For the caching problem described in Section~\ref{sys}, the optimal achievable delivery rate is lower bounded by
\begin{subequations}
\begin{align}
R^*(M)\geq&\operatorname*{max}\limits_{p\in [1: \min\{N, K\}]} \sum\limits_{s=0}^{N-p\lfloor N/p\rfloor}\sum\limits_{l=1}^{p\lfloor N/p\rfloor} \binom{N-p\lfloor N/p\rfloor}{s}\cdot \nonumber \\
&~~~~~~~ \binom{p\lfloor N/p\rfloor}{l}\frac{F_{l+s}}{\lfloor N/p\rfloor}-\frac{pM}{\lfloor N/p\rfloor}.
\end{align}
\end{subequations}
\begin{proof}
The proof will be provided in a longer version of the paper.
\end{proof}
\end{Theorem}

\section{Numerical results}\label{s:numerical}
In this section, we numerically compare the delivery rates of the proposed correlation-aware caching schemes CAUC and CACC with the lower bound and the state-of-the-art coded caching scheme from \cite{yu2017exact}, which does not take the content correlations into account. We refer to the later scheme as the correlation-ignorant coded caching scheme (CICC).

\begin{figure}[t]\label{fig:2}
\centering
\includegraphics[width=1.1\linewidth]{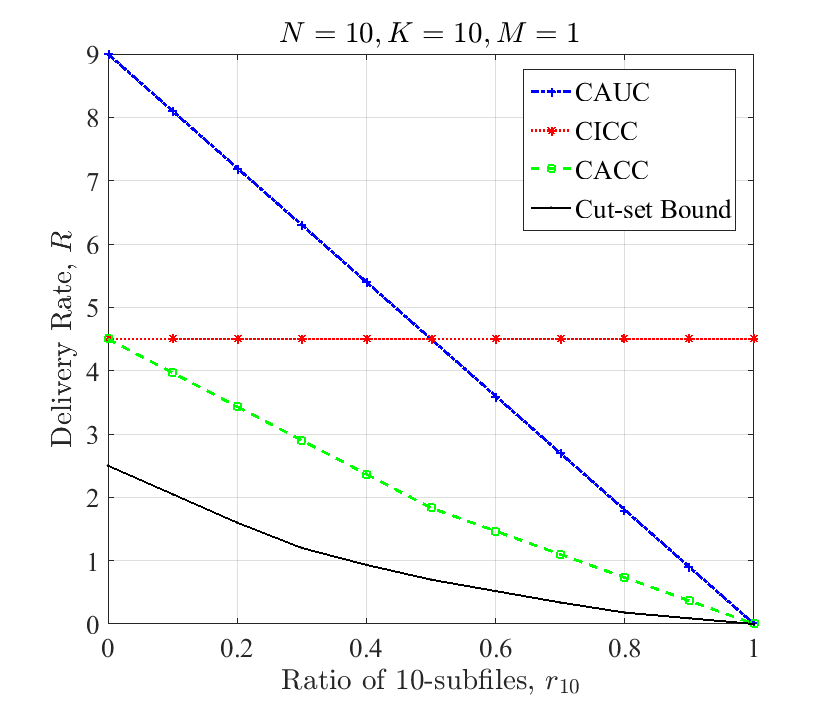}
\caption{Delivery rate ($R$) vs. ratio of $10$-subfiles ($r_{10}$), $r_2=\cdots=r_{10}=0$.}
\end{figure}

We consider $N=10$ files and $k=10$ users. Each user is equipped with a cache of size $F$ bits, i.e., $M=1$. We denote by $r_l$ the ratio of $l$-subfiles among each file, i.e., $r_l\triangleq\binom{N-1}{l-1}F_l/F$. Note that, we have $\sum_{l=1}^K r_l =1$. In Fig. \ref{fig:1}, we assume that the files have only pairwise correlations, that is, $r_3=\cdots=r_{10}=0$, and we plot the delivery rate as a function of $r_2$. Meanwhile, in Fig. \ref{fig:2}, we assume that each file consists of a private part, i.e., $1$-subfile, and a common subfile that is shared by all the files in the library, i.e., $10$-subfile, i.e., $r_2=\cdots=r_{9}=0$. We plot the delivery rate as a function of $r_{10}$. 

We observe in both figures that the delivery rate achieved by the correlation-ignorant scheme, CICC, remains the same no matter how high the ratio of common subfiles, while the delivery rates of the correlation-aware schemes, CAUC and CACC, decrease as the ratio of the common subfiles increases. Obviously, CACC achieves a lower delivery rate than both CAUC and CICC, since it benefits both from incorporating the correlations among the files as well as coded multicasting. When the ratio of the common subfiles is sufficiently large, even without coded multicasting CAUC achieves a lower delivery rate than CICC. It can also be observed that the delivery rates of correlation-aware schemes decrease faster with the percentage of common subfiles in Fig. \ref{fig:2} than in Fig. \ref{fig:1}. That is because the gain from exploiting correlation is more pronounced as the common parts are shared among more files. While there is a gap between the cut-set lower bound and the achievable delivery rate, we note that the gap is smaller in Fig \ref{fig:2}, where the level of commonness is higher. 
 
\section{Conclusions}

We have studied coded caching taking into account the available correlations among the files in the library. To capture arbitrary correlations, we assume that each file consists of a number of subfiles, each of which is shared by a different subset of files in the library, and the number of files that share a certain subfile is defined as its level of commonness. We proposed both a correlation-aware uncoded caching scheme, the optimal placement of which is proven to be caching the subfiles with the highest levels of commonness, and a correlation-aware coded caching scheme (CACC), the placement of which is optimized in terms of the achievable delivery rate. The proposed CACC scheme, or even the uncoded caching scheme when the correlation among files is strong enough, is shown to significantly outperform the best known achievable delivery rate by correlation-unaware solution in the literature.

\bibliographystyle{IEEEtran}
\bibliography{report}
\end{document}